\documentclass[%
 amsmath,amssymb,
 preprint,%
]{revtex4-1}

\usepackage{graphicx}
\usepackage{dcolumn}
\usepackage{bm}

\usepackage[utf8]{inputenc}
\usepackage[T1]{fontenc}
\usepackage{mathptmx}
\usepackage{etoolbox}

\begin{document}


\title[Plasma Telescope]{Compact Petawatt-Class Laser Wakefield Acceleration with Plasma Telescope}
\author{Xuesong Geng}
\author{Liangliang Ji}%
 \email{jill@siom.ac.cn}
 \affiliation{ 
State Key Laboratory of High Field Laser Physics and CAS Center for Excellence in Ultra-intense Laser Science, Shanghai Institute of Optics and Fine Mechanics (SIOM), Chinese Academy of Sciences (CAS), Shanghai 201800, China
}

\author{Baifei Shen}
 \email{bfshen@mail.shcnc.ac.cn}
 \affiliation{Shanghai Normal University, Shanghai 200234, China.}

\date{\today}

\begin{abstract}
The compactness of laser wakefield acceleration (LWFA) is limited by its long focal length for high power lasers, e.g., more than 10 meters for 1-peatawatt (PW) laser pulse and up to hundreds of meters for 10-100 PW lasers. The long focal length originates from the low damage threshold of the optical off-axial parabolic (OAP) mirror and consequent large spot size. We propose implementing an OAP plasma mirror (PM) to form a telescope geometry, reducing the beam size and hence constraining the focal length to meter-range for LWFA driven by lasers beyond 1PW. Three-dimensional particle-in-cell simulations are performed to characterize the reflection of a 1-PW laser by the plasma OAP and find that optimal condition is achieved within only 1-m optical length. The new method successfully generates 9GeV electron bunch in the subsequent LWFA stage with consistent acceleration gradients to that of the 1-PW laser via ordinary focusing. The proposed geometry provides a solution of compact LWFAs available for even 100-PW laser systems.
\end{abstract}

\maketitle

\section{Introduction}
Laser wakefield acceleration (LWFA) \cite{tajimaLaserElectronAccelerator1979} is able to generate high-energy electrons within a short distance, promising the ability to build compact particle accelerators. 
The state-of-the-art laser wakefield accelerator is able to accelerate electrons to the order of 10GeV within tens of centimeters using 850TW laser \cite{gonsalvesPetawattLaserGuiding2019}. 
Recent 10-100PW laser systems worldwide \cite{dansonPetawattExawattClass2019a} have enabled LWFA with higher laser powers. 
In general, higher laser power permits longer accelerating distance and higher electron energy, i.e. $\Delta E_k\sim P^{1/3}\ $ \cite{luGeneratingMultiGeVElectron2007} where $\Delta E_k$ is the energy gain and $P$ the laser power. 
The 10-100PW laser facilities will raise the energy limit to the level of 100GeV within a single stage, as shown in Fig. 1(a), which can be boosted to TeV level with the help of multistage LWFA acceleration \cite{luoMultistageCouplingLaserWakefield2018}. 
However, LWFA in the blow-out regime \cite{pukhovLaserWakeField2002a, luNonlinearTheoryRelativistic2006} requires that the laser field strength should be slightly above relativistic threshold, and the spot size will be at the order of 100$\rm{\mu m}$ for 10-100PW laser pulses, 
which requires focusing mirrors of large f-numbers, resulting in several hundred-of-meter focal length, as shown by the dashed line in Fig. 1(b) and 1(c), 
where the laser spot is assumed to be focused to a fixed field strength above relativistic threshold. 
The extremely long focal length hinders the application of high-power laser systems to multi-stage LWFA since the actual size of the LWFA system is not only the acceleration length but also the size of the whole optical system.

\begin{figure*}
\includegraphics[width=0.7\textwidth]{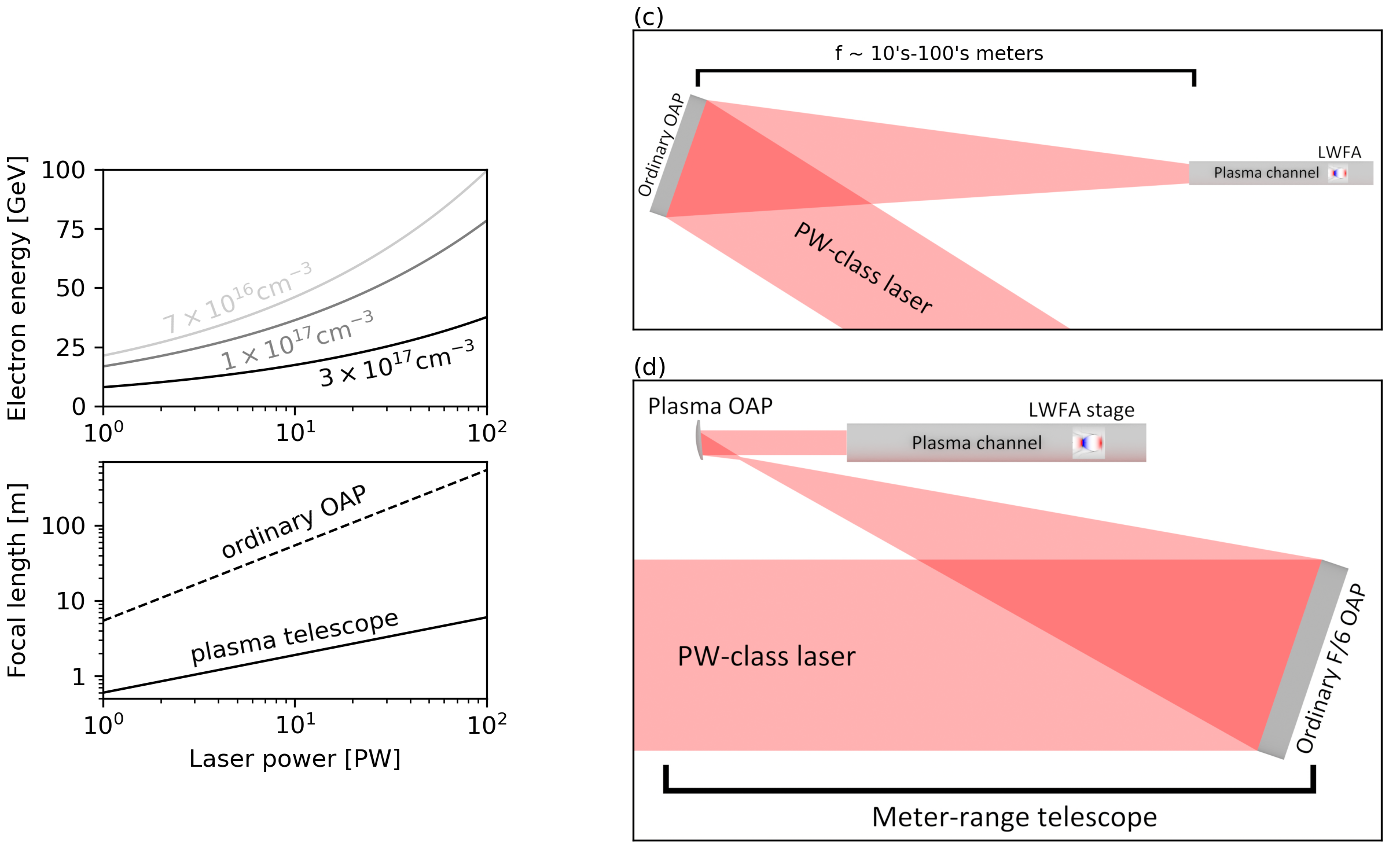}
\caption{\label{fig:fig1}
 (a) Scaling of electron energy gain in LWFA driven by PW-class laser systems. The electron energy gain depends on both the laser power and the plasma density. (b) Scaling law of focal lengths of LWFA for ordinary focusing system and the proposed plasma telescope. (c) PW-class laser focused by an OAP mirror with long focal length. (d) PW-class laser focused by plasma telescope composed of an ordinary OAP mirror with short focal length and a plasma OAP mirror behind the focal spot. It compresses the size of focusing system of PW-class lasers from 10-100 meters to a few meters.}
\end{figure*}

It is therefore crucial to shorten the optical length for LWFA if the system is to remain compact and capable of multi-stage acceleration \cite{gonsalvesTunableLaserPlasma2011, steinkeMultistageCouplingIndependent2016}. Due to the damage threshold of the amplification media of the laser system, the diameter of the laser spot before the final off-axial parabolic (OAP) mirror is large for high power lasers, prohibiting the reduction of the focal length.  For 1PW laser, the optimized acceleration length is about 20cm, while the focal length is at the order of 10m, as shown in Fig. 1(b).

Plasma, as a media free of damage threshold, provides a promising approach to manipulate high power lasers. Plasma optics provides multiple tools to control intense laser pulses, e.g., reflection of laser pulses of ultra-high intensities \cite{thauryPlasmaMirrorsUltrahighintensity2007, taphuocAllopticalComptonGammaray2012} and improving the beam contrast for laser-solid interactions \cite{thauryPlasmaMirrorsUltrahighintensity2007} by plasma mirror (PM), compression intense laser pulses by plasma gratings \cite{edwardsPlasmaTransmissionGratings2022,suntsovFemtosecondLaserInduced2009,wuLaserCompressionFastextending2022} and plasma lenses for high power lasers \cite{palastroPlasmaLensesUltrashort2015,zengPlasmaLensesRelativistic2020}.

Here, we propose using plasma telescope to transform a tightly focused pulse to a quasi-plane-wave beam of spot size suitable for LWFA. By using a telescope system, the size of the focusing system can be reduced to a few meters even for 100PW lasers, which is one order of magnitude smaller than the ordinary focusing system. 
As shown in Fig. 1(d), plasma telescope is basically composed of an ordinary OAP mirror of small f-number and an OAP plasma mirror that reflects intense laser focused by the first OAP. Curved PMs have been experimentally employed to focus intense laser to higher intensities \cite{nakatsutsumiFastFocusingShortpulse2010,arikawaUltrahighcontrastKilojouleclassPetawatt2016,wilsonEllipsoidalPlasmaMirror2016,nakatsutsumiSelfgeneratedSurfaceMagnetic2018,vincentiAchievingExtremeLight2019,quereReflectingPetawattLasers2021}, which is a demonstration of the practicality of the focusing ability of the curved PMs. By using the telescope system, the jittering of laser spot on the target, e.g., the plasma channel for LWFA, can be better stabilized than the hundreds-of-meter-long ordinary focusing system since the distance between the channel and the OAP-PM is much shorter than the ordinary focusing system.

In the following, we carry out numerical validation of the reflection of 1PW laser pulse by an OAP-PM via 3-dimesional (3D) particle-in-cell (PIC) simulations. The laser wavefront, reflection efficiency, presence of preplasma and high-order harmonics are investigated in the 3D simulations. The reflected pulse is then utilized as the driving pulse of the subsequent LWFA stage to qualify its availability, which is simulated in the quasi-cylindrical coordinate. The laser pulse reflected by plasma telescope shows consistent acceleration gradient as compared to the pulse focused by ordinary OAP, where no significant modification is induced by laser-plasma interaction, demonstrating 9GeV electron acceleration with 1PW laser power in a 1m optical length. The scenario can be extended to the 10-100PW lasers since the field strengths during reflection and the interaction geometry are almost the same.

\section{Reflection by OAP-PM}
In the following simulations, we assume the field strength of the circularly polarized driving pulse at LWFA stage is about $a_0 = 4$, i.e. $a_0 \approx 2.83$ in y and z directions for laser propagating along x, 
and peak power of 1PW, where $a_0 = eE_0/mc\omega $ is the normalized field strength with $e$ the electron charge, $E_0$ the peak electric field amplitude, $m$ the electron mass, $c$ the speed of light and $\omega$ the angular frequency of the laser. 
In order to achieve short focal lengths, the f-number of the first ordinary OAP used to focus the incident laser is fixed to $f_\# \approx 6$ which will generate a laser spot of $w_0 = 4\rm{\mu m}$ for 800nm lasers at the first focus, as shown in Fig. 2(a). 
The f-number of the first OAP mirror is flexible by adjusting the subsequent OAP-PM, as long as the combined plasma telescope transforms the incident laser to desired spot size. 
In order to get reflected pulse of $a_0 = 4$, the OAP-PM is placed behind the laser focus at $z_0$, 
where the field strength decreases to around $a_0 = 4$ and spot size increases to $w_0 \approx 43\rm{\mu m}$, corresponding to curvature radius of the wavefront of $R \approx 680\rm{\mu m}$. 
To transform the focused laser to a nearly-plane-wave for later LWFA, the distance between the laser focus and the OAP-PM satisfies $z_0 \approx f \approx 675\rm{\mu m}$, where the curvature radius of the OAP-PM is approximately twice of that of the incident wavefront, 
which guarantees the transformation of curved wavefront to flat wavefront. 
The chosen off-axial angle is $\theta = 0.15$ rad for computational efficiency consideration, which will be discussed in method section. As a result, the f-number of the OAP-PM is about $f_\# \approx 10.6$.

\begin{figure}
\includegraphics[width=0.7\textwidth]{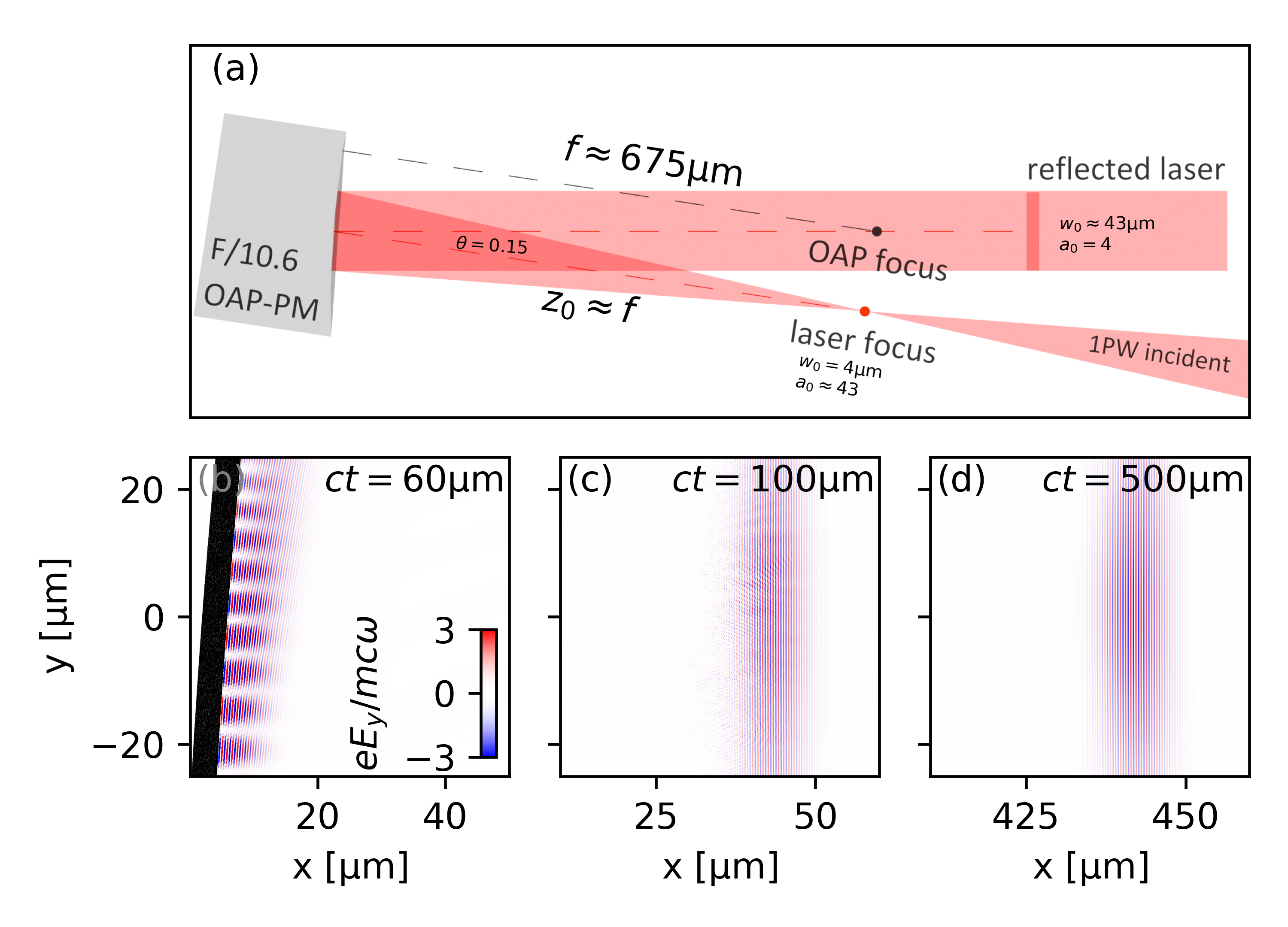}
\caption{\label{fig:fig2}
  (a) Geometry of reflection of the incident pulse by OAP-PM. (b-d) Electric fields (red and blue) and PM density (black) in the xy-plane at $ct = 60\rm{\mu m}$, $ct = 100\rm{\mu m}$ and $ct = 500\rm{\mu m}$.}
\end{figure}

The reflection and propagation of the pulse is shown in Fig. 2(b-d). During the reflection, as shown in Fig. 2(b), the field strength is periodically modulated along the y-direction by the interference between the reflected and incident pulse due to the extra optical path induced by the curved surface. On the other hand, no visible distortion of the PM is observed. 
After the pulse is reflected and the simulation window starts to move, slight modifications and noises induced by laser-plasma interaction can be observed in the behind of the pulse in Fig. 2(c), which become absent after propagation of $ct = 500\rm{\mu m}$ shown in Fig. 2(d). 
It should be noted that the density distribution and laser pulse are rotated by the off-axial angle of $\theta = 0.15$ so that the reflected pulse propagates along the x-direction. 
The reflected pulse and the transverse profile are shown in Fig. 3(a-b) and the modification to the pulse is quantified in Fig. 3(c) in terms of transverse profile of the electric fields and the transverse phase relative to the pulse center. 
The relatively flat phase curve indicates that the wavefront is nearly planar. But the lowered field profile (solid-red/-blue) indicates that a part of the pulse energy is lost after the reflection, which is 83.6\% of the incident pulse, i.e., 24.0J, due to the absorption and heating of the PM electrons.

However, in more realistic situations, the pre-pulse or pedestals of the laser pulse may produce preplasma on the front surface of the PM, which will influence the energy absorption of PM from the laser pulse.
Thus, an exponentially distributed preplasma of $\exp(-x'/l_{\rm{pre}})$ is added to the PM surface, where $x’$ is the coordinate vertical to the rotated OAP-PM and $l_{\rm{pre}}$ is the scale of the preplasma \cite{vincentiAchievingExtremeLight2019}. 
In our modelling, we choose $l_{\rm{pre}} = 0.1\rm{\mu m}$, which is a typical situation that can be realized by adjusting the laser pre-pulse. 
We notice that the reflectivity can be boosted to 93.4\% at the presence of preplasma of $l_{\rm{pre}} = 0.1\rm{\mu m}$, but at the expense of modified pulse profile, as shown in Fig. 3(d-f), where the squeezed transverse profile and convex phase indicate that the laser is more focused than that without preplasma. 
It is because the presence of preplasma amplifies the denting of the PM \cite{vincentiOpticalPropertiesRelativistic2014} which further shortens the focal length. 
In our modelling, larger $l_{\rm{pre}}$ will further amplify the denting effect and degrade reflectivity. 
In fact, the formation and evolution should be well controlled by adjusting the strength and delay of the pre-pulse whenever PM is utilized \cite{vincentiOpticalPropertiesRelativistic2014}. 

To mitigate the modification, we adjust the focal length of the OAP-PM to $f = 750\rm{\mu m}$ from the designed $f = 675\rm{\mu m}$. 
The transverse profile and the relative phase get recovered, as shown in Fig. 3(i), where the pulse profile is closer to the expected Gaussian pulse profile (dashed-gray line) and the laser phase is as flat as Fig. 3(c), 
indicating that the wavefront is almost planar and is suitable for the LWFA stage. 
As a result, by increasing the focal length in the presence of preplasma, the OAP-PM is able to reflect the incident laser pulse with high reflectivity without significant distortion of the wavefront.

\begin{figure}
\includegraphics[width=0.7\textwidth]{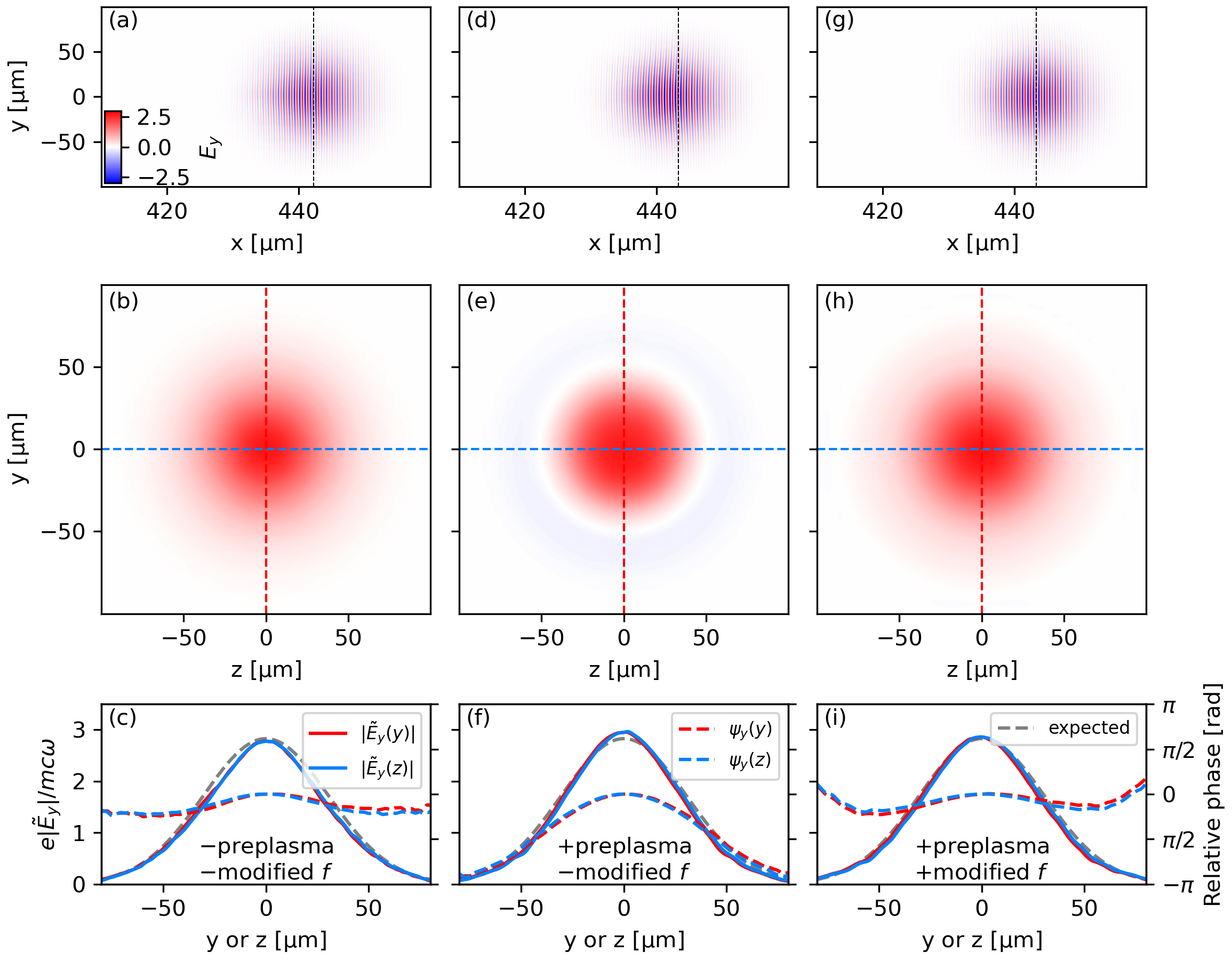}
\caption{\label{fig:fig3}
  (a) The reflected pulse at $ct = 500\rm{\mu m}$ in xy-plane for $f = 675\rm{\mu m}$ without preplasma. 
  (b) The Ey  field in the yz-plane sliced at the black-dashed lines in (a-c). 
  (c) The profile of Ey at the red-/blue-dashed lines in (d-e) (red-/blue-solid) and the corresponding phases (red-/blue-dashed) relative to the pulse center. The gray-dashed lines are the expected Ey profile without energy loss. 
  (d-f) The results of $f = 675\rm{\mu m}$ with preplasma. 
  (g-i) The results of $f = 750\rm{\mu m}$ with preplasma.}
\end{figure}

When the incident laser is linearly polarized (LP), the reflectivity, however, drops to around 73\% for both s- and p-polarizations in our modelling due to stronger ponderomotive oscillation of electrons in the LP laser fields than in the CP laser fields. 
Therefore, focusing of LP PW-class lasers with the proposed plasma telescope comes with energy loss of about 30\%, which becomes a trade-off between long focal lengths and lowered laser energies when reflecting LP lasers.
During the laser-plasma interaction, high-order harmonics could also be generated via the relativistic oscillating mirror mechanism \cite{teubnerHighorderHarmonicsLaserirradiated2009,ghimireHighharmonicGenerationSolids2019}. 
The normalized field strengths of the second and third order harmonics are $a_0(2\omega) \approx 0.25$ and $a_0(3\omega) \approx 0.15$ in our modelling, which are negligible compared to the main pulse of $a_0(\omega) \approx 4$ in the context of LWFA. 
This is because the high-order harmonics efficiency is suppressed for CP lasers \cite{easterAngularEmissionPolarization2013}. 
However, for s-/p-polarized lasers, the strength of the 3rd order harmonic can reach $a_0(3\omega) \approx 1.2$, which could potentially affect LWFA stage. 
Thus, the role of HHG in LWFA remains to be investigated for LP lasers. In the following LWFA stage, only CP laser is considered.

\section{Acceleration by reflected pulse}
Plasma channel can guide laser pulse without changing the spot size when the laser spot size matches the channel \cite{durfeeLightPipeHigh1993}, which is an effective guiding method for long-distance LWFA  \cite{gonsalvesPetawattLaserGuiding2019,leemansGeVElectronBeams2006}. 
The density profile of the channel is expressed as $n_e(r) = n_0 + r^2/\pi r_e w_m^4$ where $n_0$ is the central electron density, $r_e \approx 2.8\times10^{-15}$m the classical electron radius and $w_m$ the matched spot size.

For demonstration of the acceleration capability of the reflected laser pulse, we carry out LWFA simulation for both the reflected pulse and the pulse injected from simulation boundary, i.e., without reflection by PM. 
The laser pulses are injected into the plasma channel with central density of $n_0 \approx 2.4\times10^{17}\rm{cm^{-3}}$ and matched spot size of $w_m \approx 43\rm{\mu m}$ for $w_m = w_0$. 
The LWFA stage is simulated via FBPIC \cite{leheSpectralQuasicylindricalDispersionfree2016a} in r-z coordinates with the reflected pulse as the initial condition. Simulation details are shown in the method section. 

The accelerating field $E_z$ and bubble structures are compared in Fig. 4(a-c). One can see that the strengths of the stimulated accelerating fields $E_z$ are almost identical for the reflected and injected pulses, 
which remains true in the long-term evolution shown in Fig. 4(d) and 4(e), where the bubble and injected electron are represented by the variation of the on-axis acceleration field $E_z$ and the on-axis electron density. 
The yellow trajectories indicate the electrons of high density, i.e., the tail of the bubble and the injected electron bunch. It can be inferred from the electron bunch trajectory and the evolution of $E_z$ that the electrons experience similar acceleration gradients. 
The consequent electron energy evolution and final spectrum are shown in Fig. 4(f) and 4(g). It can be seen that the electrons injected at similar positions are accelerated to similar energies despite the different bubble evolution. 
For example, electrons injected at $ct \approx 60$mm are accelerated to $E_k \approx 6$GeV and $ct \approx 40$mm to $E_k \approx 8$GeV in both cases, which can be inferred from the gray lines in Fig. 4(f) and 4(g).

\begin{figure}
\includegraphics[width=0.7\textwidth]{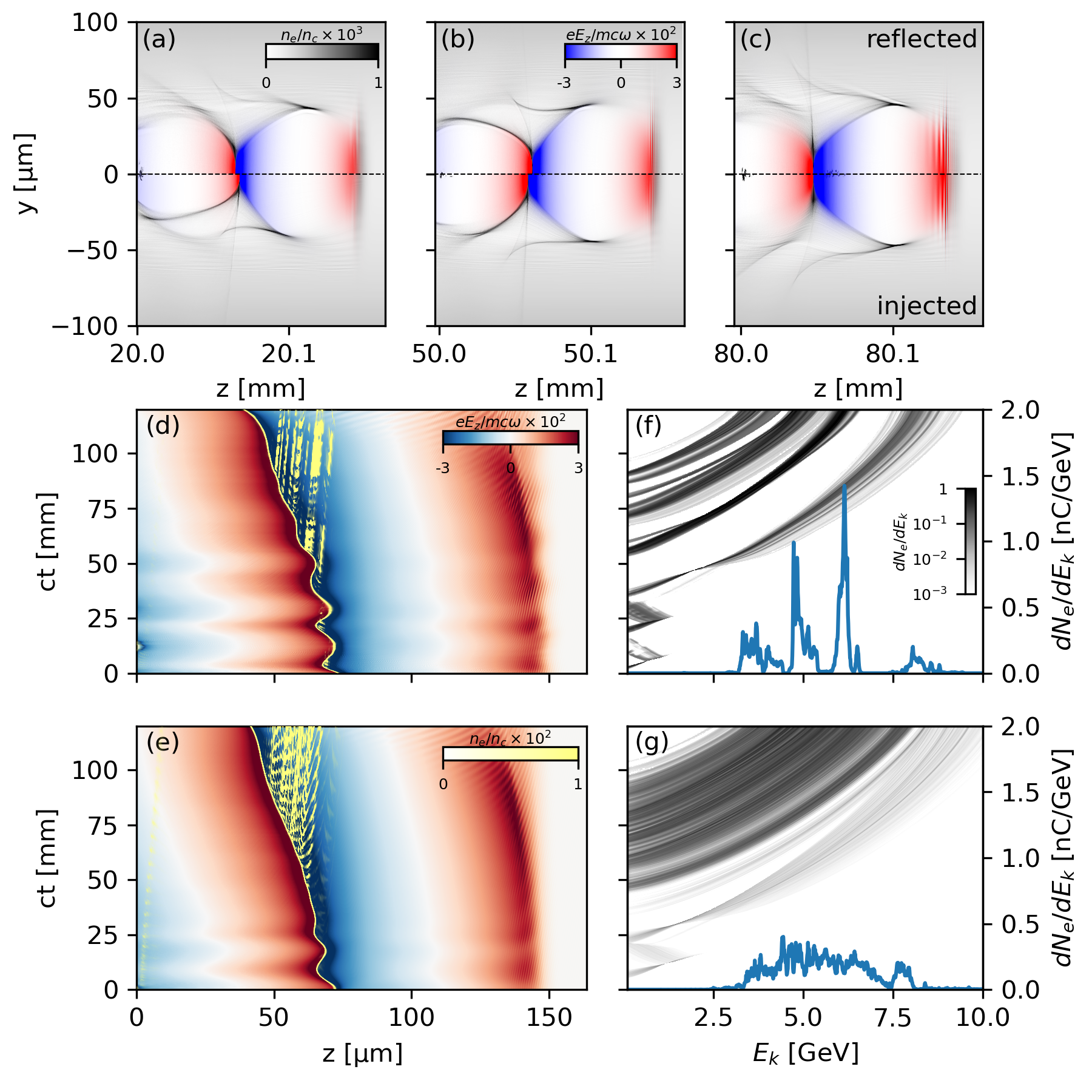}
\caption{\label{fig:fig4}
  (a-c) Accelerating field $E_z$ (red and blue) and electron density (gray) at (a) $z \approx 20$mm, (b) $z \approx 50$mm and (c) $z \approx 80$mm of the pulse reflected by plasma OAP (up) and the pulse injected from boundary (down). 
  (d-e) Variation of the on-axis accelerating field $E_z$ in the simulation window (red and blue) and the on-axis electron density (yellow) for (d) the reflected and (e) the injected pulse. 
  (f-g) Evolution of the electron energy spectrum (gray) and the final spectrum (blue) for (f) the reflected and (g) the injected pulse. }
\end{figure}

As for electron bunch emittance, both the reflected pulse and injected pulse generate well-collimated electrons as indicated by the curves in Fig. 5. The emittance is calculated by $\varepsilon_x=\sqrt{\langle x^2\rangle\langle \theta_x^2\rangle - \langle x\theta_x\rangle^2}$, 
where $\langle\cdot\rangle$ denotes the standard deviation and $\theta_x = \tan^{-1}(p_x/p_z)$, and $\varepsilon_y$ is calculated in the same way. 
In other words, for bunch size of 5$\rm{\mu m}$ the angular divergence is about 0.2mrad when $\varepsilon_x \approx \varepsilon_y \approx 1\rm{\mu m \cdot mrad}$. 
In both situations the evolutions of the bunch emittance are almost identical after the electrons are significantly injected, i.e., 
after $ct = 50$mm, except that the bunch emittance driven by the reflected pulse is slightly higher than the injected pulse due to much higher bunch charge.

\begin{figure}
\includegraphics[width=0.5\textwidth]{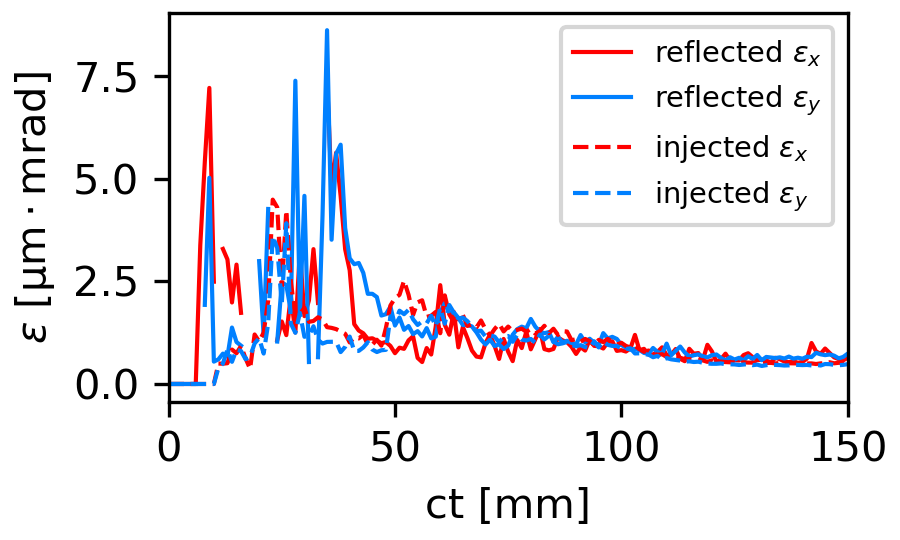}
\caption{\label{fig:fig5}
  Electron bunch emittance $\varepsilon_x$ (red) and $\varepsilon_y$ (blue) driven by the pulse reflected by plasma OAP (solid) and the pulse injected from boundary (dashed). Electrons below 1GeV are excluded.}
\end{figure}

However, since LWFA in the blowout regime is highly nonlinear \cite{pukhovLaserWakeField2002a,luNonlinearTheoryRelativistic2006,esareyPhysicsLaserdrivenPlasmabased2009} and is sensitive to initial conditions, any modification of the driving laser could induce disparate results, especially for long-distance acceleration \cite{pukhovLaserWakeField2002a,wangQuasimonoenergeticLaserplasmaAcceleration2013,kimStableMultiGeVElectron2017}. The shot-to-shot instability could be even more significant than the distortion induced by reflection. In other words, considering the similar acceleration gradient in our modelling, reflection by OAP-PM will not induce more significant instability than the shot-to-shot instability of the laser system itself.

Therefore, slight mismatch between the evolution of the bubble structures can be observed in Fig. 4(a) and 4(b) where the bubble generated by the reflected pulse is larger than the injected pulse in Fig. 4(a) and smaller in Fig. 4(b). The inconsistent bubble oscillation during pulse propagation results in different self-injection moments, as indicated by the start of the yellow slashes in Fig. 4(d) and 4(e). The reflected pulse induces longer bubble oscillation time whereas the injected pulse produces continuously expanding bubble after a few oscillations. Therefore, the reflected pulse induces discontinuous self-injection, but the electrons driven by the injected pulse are continuously injected into the end of the bubble. This difference results in disparate electron bunches as shown by the variation of energy spectrums in Fig. 4(f-g). The discontinuous injection produces several energy spikes with significantly higher beam charge but lower cutoff energy at $E_k \approx 9$GeV since a few electrons are accelerated over 10GeV by the injected pulse in Fig. 4(g) as indicated by the gray lines. 

\section{Discussion}
The proposed plasma telescope aims to reduce the focal lengths of 1-100PW laser systems when driving LWFA since higher laser power generates higher electron energies, which is shown in Fig. 1(a) according to the scaling law in \cite{luGeneratingMultiGeVElectron2007}. When using ordinary OAP focusing geometry, for specific OAP damage threshold and specific focal intensity, 
the required spot size onto the OAP scales with $w_0 \sim \sqrt{P}$ and the f-number scales with $f_\# \sim \sqrt{P}$ resulting in linear scaling of focal length with $f \sim P$, as shown by the black-dashed line in Fig. 1(b). It means generating 100GeV electron bunch requires hundreds of meters of focal length. 
By using the proposed plasma telescope, the f-number of the OAP can be fixed to, for example, $f_\# \approx 6$ in all scenarios in our modelling, 
and the scaling law becomes $f \sim \sqrt{P}$, as shown by the black-solid line in Fig. 1(b), which reduces the focal length by 2 orders when generating 100GeV electrons. 

In terms of laser pointing stability, since the plasma channel can be situated much closer to the OAP-PM, typically just a few centimeters away, the positional jittering can be effectively managed. 
In contrast, conventional systems place the plasma channel tens to hundreds of meters away from the focusing mirror. 
For instance, a 1$\rm{\mu rad}$ angular jittering of the first OAP mirror results in only about 10$\rm{\mu m}$ of positional jittering when the plasma channel is 1cm away from the OAP-PM, as illustrated in Fig. 6. The positional jittering can be further reduced to 1$\rm{\mu m}$ if the plasma channel can be placed 1mm away from the OAP-PM.
However, in conventional systems, the same angular jittering would cause positional jittering of tens to hundreds of micrometers at the plasma channel due to the longer focal length that extends up to hundreds of meters for higher power lasers. 
The plasma telescope thus alleviates the stringent requirements on the pointing stability of the OAP mirror at the plasma channel entrance for high-power laser systems. It provides a more compact and efficient solution, enhancing the overall performance and precision of laser systems.

\begin{figure}
\includegraphics[width=0.8\textwidth]{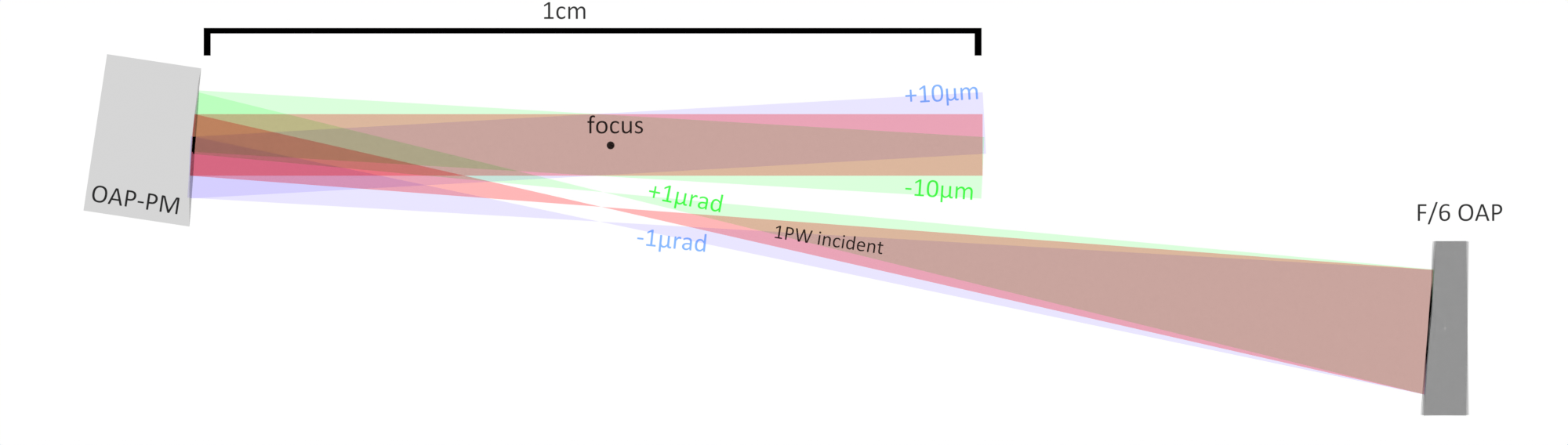}
\caption{\label{fig:fig6}
  Sketch of the positional jittering induced by angular jittering of the first OAP mirror. 1$\rm{\mu rad}$ of angular jittering will result in about 10$\rm{\mu m}$ of positional jittering at 1cm away from the OAP-PM.}
\end{figure}

For manufacturing considerations, in the investigated geometry, the microscopic OAP-PM can be manufactured via 3D-printing technique \cite{gaoLargescaleNanoshapingUltrasmooth2014} or rotating liquid that forms parabolic surface \cite{hicksonLARGEASTRONOMICALLIQUID1993}. On the other hand, the OAP-PM can be replaced by an ellipsoidal PM (EPM) \cite{nakatsutsumiFastFocusingShortpulse2010,wilsonEllipsoidalPlasmaMirror2016}.
The foci of the ellipsoid form a focus-to-focus imaging system where the spot size is magnified by $\beta/\alpha$ where $\beta$ and $\alpha$ are the distance of the foci to the reflection point of EPM \cite{stavroudisConfocalProlateSpheroids1992}. The EPM can be manufactured in macroscopic scale by tuning $\alpha$, $\beta$ and the ellipticity of the EPM.

\section{Conclusion}
The proposed OAP-PM effectively transforms a 1PW laser pulse focused by a short-focal-length OAP to a pulse with long Rayleigh length, forming a plasma telescope. The pulse reflected by OAP-PM successfully generates 9GeV electron bunch in the subsequent LWFA stage. Though slight modifications are introduced to the reflected pulse, the acceleration gradient and bunch emittance are similar to the ordinary focusing system. The proposed method essentially provides a new option to reduce the focal lengths of 1-100PW laser systems when large spot size is required like LWFA. Compact LWFA based on PW-class laser system paves way for multi-stage acceleration towards TeV electrons.

\section{Methods}
Considering the laser-plasma interaction in strong fields, the reflection of tightly focused laser pulse by OAP-PM is simulated via PIC method in 3D space, which is carried out using the EPOCH code \cite{arberContemporaryParticleincellApproach2015}. Then the LWFA stage is simulated in quasi-cylindrical coordinate via the FBPIC code \cite{leheSpectralQuasicylindricalDispersionfree2016a} based on the reflected pulse from the reflection stage, which significantly reduces the simulation time and makes it possible to simulate the 15-centimeter-long acceleration stage.

The 3-D simulations are carried out in the $80\rm{\mu m}\times200\rm{\mu m}\times200\rm{\mu m}$ box with cell size of $0.05\lambda\times0.2\lambda\times0.2\lambda$. 
The OAP-PM is placed at the left boundary of the simulation box with 8 macro-electrons and 1 macro-proton in each cell with density of $20n_c$ where $n_c$ is the critical plasma density. 
For off-axial angle $\theta$, the parabolic PM is rotated by $\theta$ along the z-axis, so that the reflected laser pulse propagates along the x-axis, as shown in Fig. 1. 

Considering the simulation efficiency, the electron density of the OAP-PM is set to $20n_c \approx 3.48\times10^{22}\rm{cm^{-3}}$ and its off-axial angle $\theta = 0.15$rad. The reflectivity could benefit from higher plasma density in the modelling, but it will require higher spatial resolution to resolve smaller plasma wavelength. Larger off-axial angles do not change the interaction picture but may need high resolution along transverse directions.

The reflected pulse in the 3-D simulation is then transformed into the cylindrical coordinate via azimuthal Fourier decomposition \cite{lifschitzParticleinCellModellingLaser2009}. First, the electromagnetic field components $F(x, y, z)$ in the Cartesian grid is converted to $F(z, r, \theta)$. The latter can be expressed by the summation of the Fourier components
\begin{equation}
    F(z,r,\theta) = \sum_m{F^{(m)} (z,r) e^{im\theta} },
\end{equation}
where $F^{(m)}(z,r)$ is the Fourier components and m is the mode number. The Fourier components are calculated via Fourier transformation
\begin{equation}
    F^{(m)}(z,r) = \frac{1}{2\pi} \int_0^{2\pi}F(z,r,\theta) e^{im\theta} d\theta.
\end{equation}
Such conversion of fields is possible when the symmetry of fields can be resolved by azimuthal Fourier expansion of $e^{im\theta}$, which is accurate when the fields are highly symmetric along the $\theta$-axis. In our simulation, 3 modes is sufficient to model the reflected pulse.
The quasi-cylindrical simulation is carried out in the $4096\times800$ window with cell size of $0.05\lambda\times0.2\lambda$ in z and r directions. Each cell is filled with $1\times1\times12$ macro-particles in z, r and $\theta$ directions. The above calculated Fourier components are loaded into the simulation window as the initial conditions.

\begin{acknowledgments}
The authors acknowledge insightful discussions with Prof. A. Pukhov and Prof. I. Kostyukov.
The work is supported by the China Postdoctoral Science Foundation (2022M713258), the Shanghai Science and Technology Development Foundation (22YF1455100), CAS Project for Young Scientists in Basic Research (YSBR060), National Natural Science Foundation of China (11935008), National Key Research and Development Program of China (2022YFE0204800), and the International Partnership Program of Chinese Academy of Sciences (181231KYSB20200040).
\end{acknowledgments}

\section*{Data Availability Statement}

The data that support the findings of this study are available from the
corresponding author upon reasonable request.

\bibliography{main}

\end{document}